\documentclass[pra,aps,showpacs,superscriptaddress,twocolumn]{revtex4}

\usepackage{amsmath}
\usepackage{bm}
\usepackage{graphicx}

\begin{document}

\title{Hysteresis in quantized vortex shedding}

\author{Tsuyoshi Kadokura}
\affiliation{Department of Engineering Science, University of
Electro-Communications, Tokyo 182-8585, Japan}

\author{Jun Yoshida}
\affiliation{Department of Engineering Science, University of
Electro-Communications, Tokyo 182-8585, Japan}

\author{Hiroki Saito}
\affiliation{Department of Engineering Science, University of
Electro-Communications, Tokyo 182-8585, Japan}

\date{\today}

\begin{abstract}
It is shown using numerical simulations that flow patterns around an
obstacle potential moving in a superfluid exhibit hysteresis.
In a certain velocity region, there is a bistability between stationary
laminar flow and periodic vortex shedding.
The bistability exists in two and three dimensional systems.
\end{abstract}


\maketitle

\section{Introduction}

The dynamics of fluids can exhibit hysteresis.
For example, a flag-like object shows bistability between flapping and
nonflapping states~\cite{Zhang,Shelley}.
Hysteresis also exists in vortex shedding dynamics behind rigid
objects, such as a vibrating cylinder~\cite{Bishop}, a multiple cylinder
arrangement~\cite{Zdra}, a long cylinder in a three-dimensional
flow~\cite{Williamson}, and a rod in a soap film~\cite{Horvath}.
In these experiments, the transitions between laminar flow and vortex
shedding states occur in a hysteretic manner as a function of the Reynolds
number.
It is known that the Taylor--Couette flow also exhibits
hysteresis~\cite{Benjamin}.
In superfluids, hysteresis has been observed in rotating toroidal
systems~\cite{Kojima,Eckel}.

In this paper, we consider the transition between a laminar flow state and
a quantized vortex shedding state around an obstacle moving in a
Bose--Einstein condensate (BEC).
In a superfluid, the velocity field around an obstacle is irrotational
below the critical velocity.
When the velocity of the obstacle exceeds the critical velocity, quantized
vortices are created and released behind the obstacle, as observed in a
trapped BEC stirred by an optical potential~\cite{Raman,Onofrio,Neely}.
The critical velocity for vortex creation and the dynamics of quantized
vortex shedding in superfluids have been studied theoretically by many
researchers~\cite{Frisch,Nore93,Jackson,Josserand,Huepe,Sties,Nore00,
Winiecki,Rica,Aftalion,Sasaki,Aioi,Saito,Pinsker,Stagg}.

The purpose of the present paper is to show that superfluids undergo
hysteretic changes between stationary laminar flow and periodic
shedding of quantized vortices.
Consider an obstacle with gradually increasing velocity; on reaching the
critical velocity $V_{c1}$, periodic vortex shedding starts.
Now consider an obstacle with gradually decreasing velocity from above
$V_{c1}$; the vortex shedding stops at a velocity $V_{c2}$.
We show that there is a bistability between these flow patterns, i.e.,
$V_{c2} < V_{c1}$.
Although hysteretic vortex shedding under a moving potential was reported
in Ref.~\cite{Aioi}, the mechanism has not been studied in detail.
In the present paper, we show that the hysteretic behaviors are due to the
fact that released vortices enhance the flow velocity around the obstacle
and induce subsequent vortex creation.
We show that the hysteretic behavior is observed for a circular obstacle
moving in a two-dimensional (2D) superfluid and a spherical obstacle
moving in a three-dimensional (3D) superfluid.

This paper is organized as follows.
Section~\ref{s:formulation} formulates the problem and describes the
numerical method.
The hysteretic dynamics are studied for a 2D system in Sec.~\ref{s:2D} and
for a 3D system in Sec.~\ref{s:3D}.
Conclusions are given in Sec.~\ref{s:conc}.

\section{Formulation of the problem}
\label{s:formulation}

We study the dynamics of a BEC at zero temperature using mean-field
theory.
The system is described by the Gross--Pitaevskii (GP) equation,
\begin{equation} \label{GP}
i \hbar \frac{\partial \Psi}{\partial t} = -\frac{\hbar^2}{2m} \nabla^2
\Psi + U(\bm{r}, t) \psi + \frac{4\pi\hbar^2 a}{m} |\Psi|^2 \Psi,
\end{equation}
where $\Psi(\bm{r}, t)$ is the macroscopic wave function, $m$ is the
atomic mass, $U(\bm{r}, t)$ is an external potential, and $a$ is the
$s$-wave scattering length.
We consider situations in which a localized potential $u$ moves at a
velocity $\bm{V}(t)$, i.e., the potential $U$ has a form,
\begin{equation}
U(\bm{r}, t) = u(\bm{r} - \int^t \bm{V}(t') dt', t).
\end{equation}
We transform Eq.~(\ref{GP}) into the frame of reference of the moving
potential $U$ by substituting the unitary transformation
\begin{equation}
\Psi(\bm{r}, t) = \exp\left[-\int^t \bm{V}(t') dt' \cdot \bm{\nabla}
\right] \psi(\bm{r}, t)
\end{equation}
into Eq.~(\ref{GP}), which yields
\begin{equation} \label{GP2}
i \hbar \frac{\partial \psi}{\partial t} = -\frac{\hbar^2}{2m} \nabla^2
\psi + i \hbar \bm{V} \cdot \bm{\nabla} \psi + u(\bm{r}, t) \psi
+ \frac{4\pi\hbar^2 a}{m} |\psi|^2 \psi.
\end{equation}
In the following, the velocity vector is taken as
\begin{equation}
\bm{V}(t) = -V(t) \hat x,
\end{equation}
where $\hat x$ is the unit vector in the $x$ direction.

We consider an infinite system, in which the atomic density $|\psi|^2$ far
from the moving potential is constant $n_0$.
For the density $n_0$, the healing length $\xi$ and the sound velocity
$v_s$ are defined as
\begin{equation} \label{xi}
\xi = \frac{1}{\sqrt{4 \pi n_0 a}}, \qquad
v_s = \frac{\hbar \sqrt{4 \pi n_0 a}}{m},
\end{equation}
which determine the characteristic time scale,
\begin{equation}
\tau = \frac{\xi}{v_s}.
\end{equation}
The chemical potential for the density $n_0$ is given by
\begin{equation} \label{mu}
\mu = \frac{4\pi\hbar^2 a}{m} n_0.
\end{equation}
Normalizing Eq.~(\ref{GP2}) by the quantities in
Eqs.~(\ref{xi})--(\ref{mu}), we obtain
\begin{equation} \label{GPn}
i \frac{\partial \tilde\psi}{\partial \tilde t} = -\frac{1}{2}
\tilde\nabla^2 \tilde\psi + i \tilde{\bm{V}} \cdot \tilde{\bm{\nabla}}
\tilde\psi + \tilde u \tilde\psi + |\tilde\psi|^2 \tilde\psi,
\end{equation}
where $\tilde\psi = n_0^{-1/2} \psi$, $\tilde t = t / \tau$,
$\tilde{\bm{\nabla}} = \xi \bm{\nabla}$, $\tilde{\bm{V}} = \bm{V} / v_s$,
and $\tilde u = u / \mu$ are dimensionless quantities.
The independent parameters in Eq.~(\ref{GPn}) are only $\tilde{\bm{V}}$
and $\tilde u$.

We numerically solve Eq.~(\ref{GPn}) using the pseudo-spectral
method~\cite{Recipes}.
The initial state is the stationary state of Eq.~(\ref{GPn}) for a
velocity $V$ below the critical velocity $V_{c1}$ for vortex nucleation,
which is prepared by the imaginary-time propagation
method~\cite{Dalfovo}.
The initial state is a stationary laminar flow and contains no vortices.
To break the exact numerical symmetry, a small random noise is added to
each mesh of the initial state.
The real-time propagation of Eq.~(\ref{GPn}) is then calculated with a
change in the velocity $V$ or the potential $u$ to trigger the vortex
creation.
The size of the space is taken to be large enough and the periodic
boundary condition imposed by the pseudo-spectral method does not affect
the dynamics around the potential.

\section{Numerical results}

\subsection{Two dimensional system}
\label{s:2D}

First, we consider a 2D space.
Typically, the size of the numerical space is taken to be $512 \xi$ in $x$
and $256 \xi$ in $y$, and is divided into a $8192 \times 4096$ mesh.
The obstacle potential is given by
\begin{equation} \label{U}
u(\bm{r}) = \left\{
\begin{array}{ll}
\infty & (\sqrt{x^2 + y^2} < R), \\
0 & (\sqrt{x^2 + y^2} > R),
\end{array} \right.
\end{equation}
where $R$ is the radius of the circular potential.
Numerically, a value that is significantly larger than the chemical
potential is used for $\infty$ in Eq.~(\ref{U}).
The following results are qualitatively the same as those for a Gaussian
potential in place of the rigid circular potential in Eq.~(\ref{U}).

\begin{figure}[tbp]
\includegraphics[width=8cm]{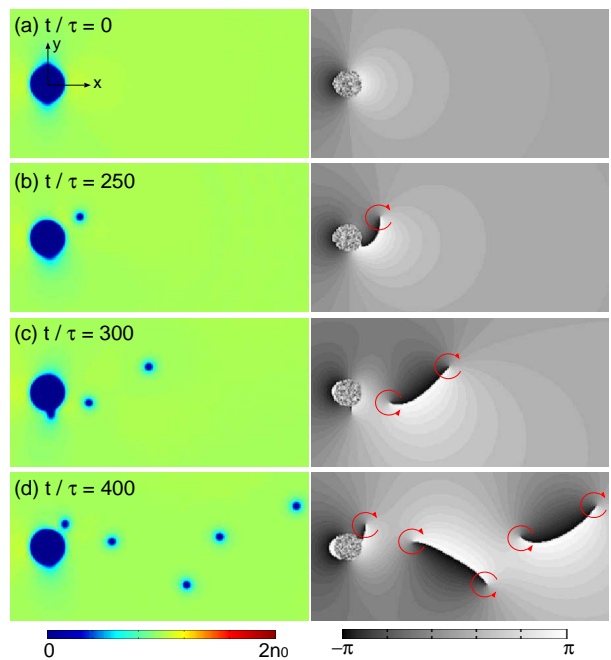}
\caption{
(color online) Time evolution of the density $|\psi|^2$ (left panels) and
phase ${\rm arg} \psi$ (right panels) profiles for $V = 0.43 v_s$ and $R =
4.1 \xi$.
To trigger the vortex shedding, the additional potential given by
Eq.~(\ref{Uadd}) is applied during $200 < t / \tau < 220$.
The arrows in the phase profiles indicate the directions in which the
quantized vortices are rotating.
The size of each panel is $80\xi \times 40\xi$.
See the Supplemental Material for a movie of the dynamics.
}
\label{f:dynamics}
\end{figure}
Figure~\ref{f:dynamics} shows the time evolution of the density $|\psi|^2$
and phase ${\rm arg} \psi$ profiles.
The initial state is the stationary state for the velocity $v = 0.43 v_s$
and radius $R = 4.1\xi$, as shown in Fig.~\ref{f:dynamics}(a).
This stationary laminar flow state is stable.
To trigger the vortex shedding, we apply an additional potential,
\begin{equation} \label{Uadd}
u_{\rm add}(\bm{r}) = \mu e^{-[x^2 + (y - R)^2] / \xi^2}
\end{equation}
during $200 < t / \tau < 220$, in addition to the circular potential in
Eq.~(\ref{U}).
This additional potential perturbs the edge of the circular potential, at
which quantized vortex creation is induced, as shown in
Fig.~\ref{f:dynamics}(b).
Subsequently, quantized vortices are periodically created one after
the other~\cite{Sasaki}, as shown in Figs.~\ref{f:dynamics}(c) and
\ref{f:dynamics}(d), even after the perturbation potential is removed at
$t = 220 \tau$ and the velocity $v = 0.43 v_s$ is smaller than the
critical velocity $V_{c1}$.
This result indicates that there are at least two stable flow patterns for
the same parameters: a stationary laminar flow and periodic vortex
shedding.

\begin{figure}[tbp]
\includegraphics[width=8cm]{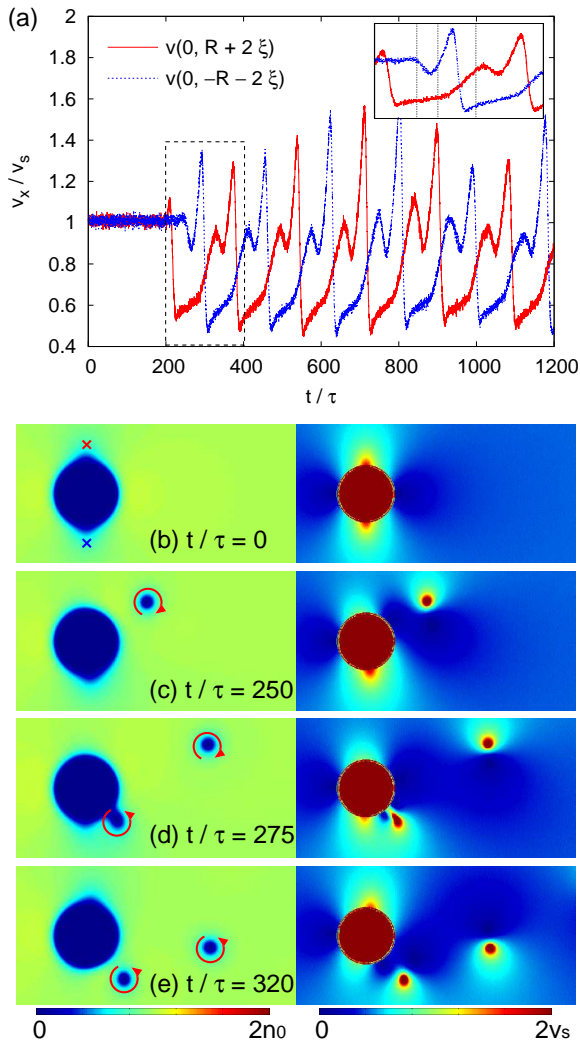}
\caption{
(color online) (a) Time evolution of the velocity $v_x$ at $(x, y) =
(0, \pm R \pm 2\xi)$ for the same parameters as those in
Fig.~\ref{f:dynamics}.
The dashed square is magnified in the inset, where the vertical lines
indicate $t / \tau = 250$, 275, and 320.
(b)--(e) Density $|\psi|^2$ profiles (left panels) and velocity $|\bm{v}|$
profiles (right panels).
The crosses in (b) indicate the positions $(x, |y|) = (0, R + 2 \xi)$ at
which the velocities are plotted in (a).
The arrows in the density profiles indicate the directions in which the
quantized vortices are rotating.
The size of each panel is $40\xi \times 20\xi$.
}
\label{f:velocity}
\end{figure}
The velocity field of the atomic flow has the form,
\begin{equation}
\bm{v}(\bm{r}, t) = \frac{\hbar}{2mi |\psi|^2} \left(
\psi \bm{\nabla} \psi^* - \psi^* \bm{\nabla} \psi \right) - \bm{V}.
\end{equation}
Figure~\ref{f:velocity}(a) shows the time evolution of the velocities
$v_x$ at $(x, |y|) = (0, R + 2 \xi)$.
These positions are indicated by the crosses in Fig.~\ref{f:velocity}(b).
For the stationary flow ($t < 200\tau$), the velocities are $v_x \simeq
v_s$.
The fluctuations around $v_s$ are due to the small numerical noises added
to the initial state.
At $t = 200 \tau$, the additional potential given by Eq.~(\ref{Uadd}) is
applied and a clockwise vortex is released from near the position $(0,
R)$.
As a consequence, $v_x(0, R + 2 \xi)$ suddenly decreases.
It can also be seen in Fig.~\ref{f:velocity}(c) that the released vortex
decreases the velocity field in the vicinity of its creation.
The clockwise vortex shedding then induces counterclockwise vortex
creation, as shown in Fig.~\ref{f:velocity}(d).
Immediately after that ($t = 275 \tau$-$320 \tau$), $v_x(0, -R - 2\xi)$
increases rapidly, which is followed by a sudden decrease due to the
shedding of another counterclockwise vortex, as shown in
Fig.~\ref{f:velocity}(d).
This periodic vortex shedding is repeated indefinitely.
The dynamics shown in Fig.~\ref{f:velocity} implies that the release of a
vortex induces the creation of a subsequent vortex, i.e., periodic vortex
shedding is taking place.

\begin{figure}[tbp]
\includegraphics[width=8cm]{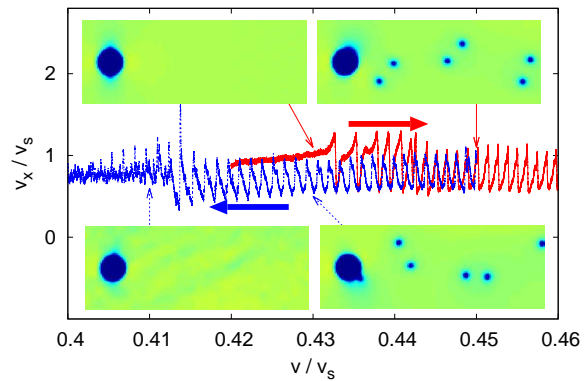}
\caption{
(color online) Time evolution of $v_x$ at $(x, y) = (0, R + 2 \xi)$ for
$R = 4.1\xi$.
The velocity is increased as $V(t) / v_s = 0.42 + 10^{-5} t / \tau$ (red,
solid line) or decreased as $V(t) / v_s = 0.45 - 10^{-5} t / \tau$
(blue, dashed line).
The insets show the density profiles at $V(t) / v_s = 0.43$ and $0.45$ for
the increase in $V(t)$, and $0.43$ and $0.41$ for the decrease in $V(t)$.
}
\label{f:hysteresis}
\end{figure}
To show the hysteresis clearly, we gradually increase and decrease the
velocity $V(t)$ around the critical velocity.
Figure~\ref{f:hysteresis} shows the time evolution of the flow velocity
$v_x$ at $(x, y) = (0, R + 2 \xi)$.
When the velocity $V(t)$ is gradually increased, the vortex shedding
starts at the critical velocity $V_{c1} \simeq 0.432 v_s$.
On the other hand, when $V(t)$ is decreased from above $V_{c1}$, the
periodic vortex shedding continues for $V(t) < V_{c1}$, eventually
stopping at the lower critical velocity $V_{c2} \simeq 0.412 v_s$.
The fluctuation in $v_x$ for $V(t) \lesssim 0.41 v_s$ is due to the
remnant disturbing waves.

\begin{figure}[tbp]
\includegraphics[width=8cm]{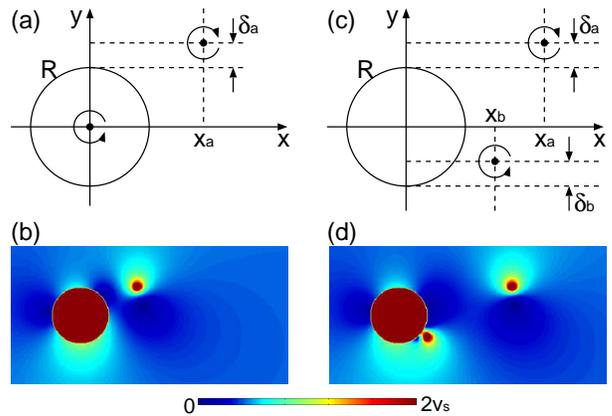}
\caption{
(color online) (a) and (c) Schematic illustrations of the point-vortex
model used to derive Eqs.~(\ref{vf1}) and (\ref{vf2}), respectively.
(b) $|\bm{v}|$ in Eq.~(\ref{vf1}) for $x_a = 2 R$ and $\delta_a = 0$.
(d) $|\bm{v}|$ in Eq.~(\ref{vf2}) for $x_a = 4 R$, $x_b = R$, $\delta_a =
0$, and $\delta_b = \xi$.
The size of each panel in (b) and (d) is $40\xi \times 20\xi$.
}
\label{f:analytic}
\end{figure}
The velocity field around a circular obstacle can be analyzed using the
point-vortex model for an inviscid incompressible fluid.
The situation in Fig.~\ref{f:velocity}(c) is modeled as in
Fig.~\ref{f:analytic}(a), where a clockwise vortex is located at $(x, y) =
(x_a, R + \delta_a)$ and the circle of radius $R$ contains a
counterclockwise vortex.
The complex velocity field in which the normal component $\bm{v} \cdot
\bm{r}$ vanishes at $r = R$ is given by
\begin{equation} \label{vf1}
v_x - i v_y = V \left( 1 - \frac{R^2}{z^2} \right)
+ \frac{\Gamma}{2\pi i} \left( -\frac{1}{z - z_a}
+ \frac{1}{z - z_a'} \right),
\end{equation}
where $z = x + i y$, $\Gamma = h / m$, $z_a = x_a + i(R + \delta_a)$, and
$z_a' = R^2 / z_a^*$~\cite{Lamb}.
The first term on the right-hand side of Eq.~(\ref{vf1}) approaches a
uniform flow $(v_x, v_y) = (V, 0)$ at infinity $|z| \rightarrow \infty$
and the second term represents a flow generated by the vortices located at
$z_a$ and the origin.
The flow velocity at $z = \pm i R$ is
\begin{equation}
v_x = 2V \mp \frac{\Gamma}{2\pi} \frac{x_a^2 + \delta_a (2R + \delta_a)}
{x_a^2 + (R \mp R + \delta_a)^2}
\end{equation}
and $v_y = 0$, which indicates that the flow velocity at $z \simeq -i R$
is enhanced by the vortices.
Thus, once a vortex is released from $z \simeq i R$, the next vortex
is created at $z \simeq -i R$, which results in the dynamics shown in
Fig.~\ref{f:velocity}(d).
The velocity field in Eq.~(\ref{vf1}) for $x_a = 2R$ and $\delta_a = 0$ is
shown in Fig.~\ref{f:analytic}(b), which is very similar to
Fig.~\ref{f:velocity}(c).

The situation in Fig.~\ref{f:velocity}(d) is modeled by
Fig.~\ref{f:analytic}(c), for which the velocity field is given by
\begin{eqnarray} \label{vf2}
v_x - i v_y & = & V \left( 1 - \frac{R^2}{z^2} \right)
+ \frac{\Gamma}{2\pi i} \biggl( -\frac{1}{z - z_a}
+ \frac{1}{z - z_a'}
\nonumber \\
& & + \frac{1}{z - z_b} - \frac{1}{z - z_b'} \biggr),
\end{eqnarray}
where $z_b = x_b + i (-R + \delta_b)$ and $z_b' = R^2 / z_b^*$.
The flow velocity at $z = -i R$ is
\begin{equation} \label{vx2}
v_x = 2V + \frac{\Gamma}{\pi} \left[ \frac{\delta_b}{x_b^2 + \delta_b^2}
- \frac{2 R + \delta_a}{x_a^2 + (2 R + \delta_a)^2} \right].
\end{equation}
When $\delta_b$ is positive, the first term in the square bracket of
Eq.~(\ref{vx2}), i.e., the vortex at $z = z_b$, enhances the flow velocity.
The vortex released from $z \simeq -i R$ therefore induces the creation of
the subsequent vortex at $z \simeq -i R$.
The velocity field in Eq.~(\ref{vf2}) for $x_a = 4R$, $\delta_a = 0$,
$x_b = R$, and $\delta_b = \xi$ is shown in Fig.~\ref{f:analytic}(d),
which well reproduces Fig.~\ref{f:velocity}(d).
Thus, vortices shed behind an obstacle induce the creation of an
additional vortex, resulting in periodic vortex shedding, and ultimately
hysteretic mechanism that allows this behavior to continue below the
critical velocity $V_{c1}$.

\begin{figure}[tbp]
\includegraphics[width=8cm]{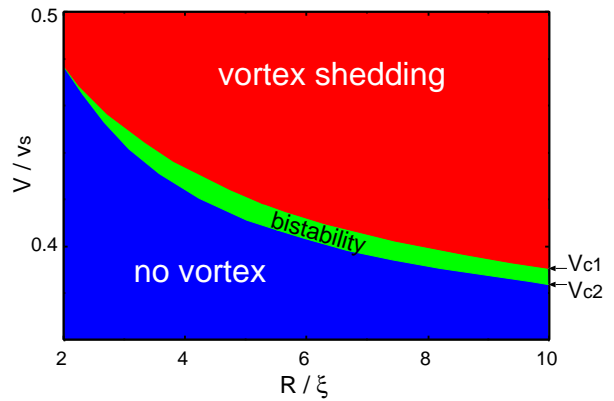}
\caption{
(color online) Parameter diagram with respect to the obstacle radius $R$
and the velocity $V$.
The vortex shedding always occurs in the ``vortex shedding'' region,
no vortices are created in the ``no vortex'' region, and hysteresis
appears in the ``bistability'' region.
The boundaries between the bistability and vortex-shedding regions and the
bistability and no-vortex regions are $V_{c1}$ and $V_{c2}$, respectively.
}
\label{f:diagram}
\end{figure}
Figure~\ref{f:diagram} shows the radius $R$ and the velocity $V$
dependence of the flow patterns.
The vortex shedding always occurs in the ``vortex shedding'' region and
vortices are never created in the ``no vortex'' region.
The ``bistability'' region lies between these two regions, in which a
stationary laminar flow is stable but periodic vortex shedding is kept
once it starts.
For $R \lesssim 2\xi$, the bistability region disappears, probably because
$\delta_b \sim R$ in Eq.~(\ref{vx2}) is small and hence the enhancement of
the successive vortex creation is less effective.
Although the bistability region may also exist for $R \gtrsim 10\xi$, it
is difficult to determine the precise value of $V_{c2}$ numerically, since
the vortex shedding dynamics are aperiodic for large $R$, and are
dependent on infinitesimal numerical noises.

\subsection{Three dimensional system}
\label{s:3D}

\begin{figure}[tbp]
\includegraphics[width=8cm]{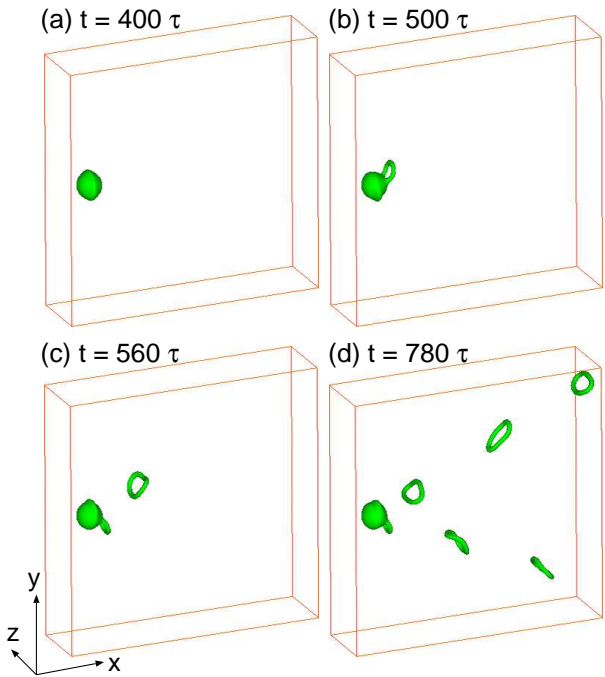}
\caption{
(color online) Isodensity surfaces for the 3D dynamics.
The Gaussian potential given by Eq.~(\ref{gaussian}) moves in the $-x$
direction at a velocity $V = 0.6 v_s$;
the dynamics are given in the frame of reference of the potential.
An additional potential given by Eq.~(\ref{add3d}) is applied during $400
\tau < t < 420 \tau$.
The size of the cuboidal frame is $126.4 \xi \times 126.4 \xi \times 31.6
\xi$.
See the Supplemental Material for a movie of the dynamics.
}
\label{f:3d}
\end{figure}
Next we examine a 3D system.
We use a Gaussian potential for the moving obstacle as
\begin{equation} \label{gaussian}
u(\bm{r}) = 100 \mu e^{-0.15 (x^2 + y^2 + z^2) / \xi^2}.
\end{equation}
A spherical rigid potential analogous to that in Eq.~(\ref{U}) gives
similar results.
We prepare the initial state of a stationary laminar flow with $V = 0.6
v_s$, which is below the critical velocity $V_{1c} \simeq 0.601$ for
vortex creation.
In order to trigger the vortex shedding, an additional potential
\begin{equation} \label{add3d}
u_{\rm add}(\bm{r}) = 0.2 \mu e^{-0.6 [x^2 + (y - d)^2 + z^2] / \xi^2}
\end{equation}
is applied during $400 \tau < t < 420 \tau$, where $d = 6.3 \xi$.
This additional potential is located at the edge of the potential $u$ in
Eq.~(\ref{gaussian}), triggering the generation of a quantized vortex
ring, as shown in Fig.~\ref{f:3d}(b).
Subsequent vortex creation is induced and periodic vortex shedding begins,
as shown in Figs.~\ref{f:3d}(c) and \ref{f:3d}(d), respectively.
This result indicates that the bistability between the stationary laminar
flow and periodic vortex shedding also exists in a 3D system.

We note that the vortex rings in a 3D system are topologically different
from the vortex pairs in a 2D system.
A vortex-antivortex pair in a 2D system corresponds to a vortex ring in a
3D system, since they propagate without changing their shapes.
By contrast, a vortex-vortex pair in a 2D system, such as that seen in
Fig.~\ref{f:velocity}(e), has no counterpart in a 3D system, since 
two vortices rotate around one another for a vortex-vortex pair.
Such rotation would tangle vortex rings in a 3D system.
It is interesting that both 2D and 3D systems exhibit bistability despite
the topological difference.

\section{Conclusions}
\label{s:conc}

We investigated the dynamics of a BEC with a moving obstacle potential,
and found bistability between stationary laminar flow and periodic vortex
shedding.
When the velocity of the obstacle is gradually increased, quantized vortex
shedding starts at the critical velocity $V_{c1}$.
On the other hand, when the velocity is gradually decreased from above
$V_{c1}$, the vortex shedding stops at a velocity $V_{c2}$.
We found that $V_{c1} > V_{c2}$ for an appropriately sized obstacle
potential (Fig.~\ref{f:hysteresis}).
For a velocity $V_{c1} > V > V_{c2}$, a stationary laminar flow is stable,
but periodic vortex shedding is maintained once it starts
(Figs.~\ref{f:dynamics} and \ref{f:velocity}).
Such hysteretic behavior originates from the fact that the vortices
released behind the obstacle enhance the velocity field around the
obstacle, inducing subsequent vortex generation (Fig.~\ref{f:analytic}).
The bistability between the stationary laminar flow and periodic vortex
shedding exists not only in 2D systems but also in 3D systems
(Fig.~\ref{f:3d}).

\begin{acknowledgments}
We thank T. Kishimoto for fruitful discussions.
This work was supported by a Grant-in-Aid for Scientific Research (No.\
26400414) and a Grant-in-Aid for Scientific Research on Innovative Areas
``Fluctuation \& Structure'' (No.\ 25103007) from the Ministry of
Education, Culture, Sports, Science and Technology of Japan.
\end{acknowledgments}

\end{document}